\documentclass[sort&compress,preprint,3p,number]{elsarticle}
\usepackage{floatflt}
\usepackage{amsmath,amssymb,xcolor}
\usepackage{hyperref}%lineno,
\usepackage{epstopdf}
\usepackage{xifthen}
\usepackage{xspace}
\allowdisplaybreaks

\renewcommand{\Im}{\operatorname{Im}}

\newcommand{\ee}{e^+e^-}
\newcommand{\citelater}[1][]{%
	{\color{red}
	\ifthenelse{\equal{#1}{}}{[\raisebox{0.1em}{\tiny \parbox[c]{3.5em}{ insert \\ citation}}]}{[#1]}}
	\xspace%
}
\begin{document}
%TODO: не забыть: изменена нумерация знаменателей так, что разрезанные знаменатели стоят в начале, т.е., поставил первые четыре в конец (они являются неприводимыми числителями).

\begin{frontmatter}

\title{
Total Born cross section of $e^+e^-$-pair production in relativistic ion collisions from differential equations.
}



\author[binp]{Roman N. Lee}
\ead{r.n.lee@inp.nsk.su}
\author[binp,nsu]{Kirill T. Mingulov}
\ead{k.t.mingulov@gmail.com}

\address[binp]{Budker Institute of Nuclear Physics, 630090, Novosibirsk, Russia}
\address[nsu]{Novosibirsk State University, 630090, Novosibirsk, Russia}

\begin{abstract}
We apply the differential equation method to the calculation of the total Born cross section of the process $Z_1Z_2\to Z_1Z_2e^+e^-$. We obtain explicit expression for the cross section exact in the relative velocity of the nuclei.
\end{abstract}

\end{frontmatter}

\section{Introduction}

Theoretical investigation of electromagnetic $\ee$ pair production in relativistic heavy-ion collisions goes back to the paper \cite{LandLif1934} where the Born cross section of the process at high energy was calculated in the leading logarithmic approximation. Racah, in his remarkable paper \cite{Racah1937}, has calculated the high-energy asymptotics of the Born cross section up to power-suppressed terms in $1/\gamma$ ($\gamma$ is a Lorentz factor of the colliding nuclei). Recently there was a certain rise of the interest to this process connected with the functioning of heavy ion colliders, like RHIC and LHC, see Ref. \cite{Baltz2008}. In particular, a great attention has been paid to the investigation of the Coulomb corrections to the cross section at high energies \cite{BaltMcL1998,IvaScSe1999,EiReScG1999,SegeWel1999,LeeMil2000}.

Speaking of the total Born cross section, the problem of its calculation is of a three-loop complexity level. Probably, this is the main reason why this quantity was not calculated exactly at arbitrary velocities of the colliding nuclei. This is in contrast to the Born cross section of pair photoproduction in the field of an ion, where the total Born cross section is known exactly for any energy of the initial photon since Refs. \cite{Racah1934a,Racah1936}. Now that we have an essential progress in the multiloop calculations, we are in position to fill this gap and to calculate the total Born cross section of $\ee$ pair production in relativistic ion collisions. 

The consideration of the present paper is based on the following approach. Using the optical theorem we express the total cross section via the sum of cut three-loop integrals. Then we apply the standard approach to multiloop calculations, based on the IBP reduction and differential equations for master integrals. The differential equations for the master integrals are first reduced to $\epsilon$-form \cite{Henn2013} using the algorithm of Ref. \cite{Lee2014}, and then solved recursively up to the required order in $\epsilon$. Thus, we obtain the total Born cross section exactly in the relative velocity $\beta$ of the colliding nuclei. Our result perfectly agrees with the celebrated result of Racah \cite{Racah1937} in the limit of large relativistic factor. At small $\beta$ we compare our result with estimate obtained in the recent paper \cite{Khriplovich2014} and find a complete disagreement. In order to find the origin of the disagreement, we perform a straightforward calculation of the low-energy asymptotics of the cross section differential with respect to the electron and positron momenta. The direct integration then reproduces our result obtained with the help of the differential equations.

\section{Born cross section for the production of $e^+e^-$ pair}

\begin{figure}
\centering
\includegraphics[width=0.7\linewidth]{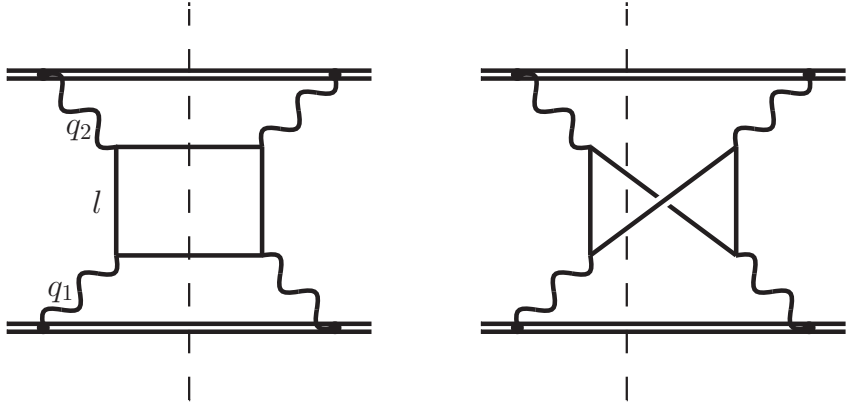}
\caption{Cut diagrams for the calculation of the total cross section of $\ee$ pair production in the collisions of relativistic nuclei. Cut thin line denotes the cut propagator $-2\pi i\delta(p^2-m^2)(\hat{p}+m)$ of the electron, cut double line denotes the cut propagator $-2\pi i\delta(2u\cdot q)$ of a heavy particle, interaction vertex with the heavy particle is $-iu^\mu$ ($u=P/M$ is a four-velocity of the heavy particle). 
}
\label{fig:cut_diagrams}
\end{figure}

Using optical theorem, the total cross section of the process $Z_1Z_2\to Z_1Z_2\ee$ can be written as
\begin{equation}
\sigma=\frac{8\Im\mathcal{A}}{\gamma\beta}\,,
\end{equation}
where $\Im\mathcal{A}$ is given by the sum of two cut diagrams depicted in Fig. \ref{fig:cut_diagrams}, $\beta$ is the relative velocity of the colliding nuclei, and $\gamma=[1-\beta^2]^{-\frac12}$ is the Lorentz factor.
Contribution of both diagrams can be expresses in terms of the scalar integrals
\begin{gather}
I(n_1,\ldots,n_{12})=\int \frac{d^dl\,d^dq_1\,d^dq_2}{(2\pi)^{3d}}\theta(q_1^0-l^0)\theta(q_2^0+l^0)\prod_{k=1}^4\Im\frac{1}{(D_k+i0)^{n_k}}\prod_{k=5}^{12}\frac{1}{(D_k+i0)^{n_k}}\,,\nonumber\\
D_1=-2 q_1\cdot u_1\,,\quad 
D_2=-2 q_2\cdot u_2\,,\quad
D_3=\left(l-q_1\right)^2-1\,,\quad
D_4=\left(l+q_2\right)^2-1\,,\nonumber\\
D_5=l^2-1\,,\quad
D_6=\left(l-q_1+q_2\right)^2-1\,,\quad
D_7=q_1^2\,,\quad 
D_8=q_2^2\,,\nonumber\\
D_9=-2 l\cdot u_1\,,\quad 
D_{10}=-2 l\cdot u_2\,,\quad 
D_{11}=-2 q_2\cdot u_1\,,\quad 
D_{12}=-2 q_1\cdot u_2\,.
\end{gather}
Here $u_1$ and $u_2$ are the four-velocities of the nuclei, so that $u_1\cdot u_2=\gamma$.

We proceed in the following way. First, we perform the IBP reduction of the cut integrals from the above topologies in $d=4-2\epsilon$. For this step we use \texttt{LiteRed}, Refs. \cite{Lee2012,Lee2013a}.
We end up with 8 master integrals
\begin{gather*}
	J_1={\scriptstyle I(1,1,1,1,0,0,0,1,0,0,0,0)},\ J_2={\scriptstyle I(1,1,1,1,0,1,0,0,0,0,0,0)},\ J_3={\scriptstyle I(1,1,1,1,0,2,0,0,0,0,0,0)},\\
	 J_4={\scriptstyle I(1,1,1,1,-1,1,0,0,0,0,0,0)},\ J_5={\scriptstyle I(1,1,1,1,0,0,1,1,0,0,0,0)},\ J_6={\scriptstyle I(1,1,1,1,1,1,0,0,0,0,0,0)},\\ 
	J_7={\scriptstyle I(1,1,1,1,0,0,1,1,-1,-1,0,0)},\ J_8={\scriptstyle I(1,1,1,1,1,1,0,0,-1,0,-1,0)}\,.
\end{gather*}

Introducing the column-vector $\mathbf{J} = (J_1,\ldots, J_{8})^T$,
we obtain the differential system
\begin{equation}\label{eq:system}
\frac{\partial }{\partial\gamma}\mathbf{J} =M(\gamma,\epsilon)\mathbf{J}\,,
\end{equation}
where $M(\gamma,\epsilon)$ is a matrix with entries being rational functions of both $\gamma$ and $\epsilon$. Passing to new variable, $x=\frac{1-\beta}{1+\beta}$, we apply the algorithm from Ref. \cite{Lee2014} to reduce the differential system \eqref{eq:system} to $\epsilon$-form \cite{Henn2013}. The differential system for the new basis $\widetilde{\mathbf{J}} = (\widetilde{J}_1,\ldots, \widetilde{J}_{8})^T$ has the form
\begin{equation}
\frac{\partial }{\partial x}\widetilde{\mathbf{J}} =\epsilon \left[\frac1{x}M_0+\frac{1}{x-1}M_1+\frac{1}{x+1}M_2\right]\widetilde{\mathbf{J}}\,,
\end{equation}

%NB: I used permutation {1, 5, 7, 2, 3, 4, 6, 8} of columns and rows in comparison with the matrix in the file.
\begin{align}
M_0&={\arraycolsep=1.4pt\tiny\left[
	\begin{array}{cccccccc}
	-1 & 0 & 0 & 0 & 0 & 0 & 0 & 0 \\
	1 & 0 & 0 & 0 & 0 & 0 & 0 & 0 \\
	0 & 1 & -1 & 0 & 0 & 0 & 0 & 0 \\
	0 & 0 & 0 & -1 & 0 & 1 & 0 & 0 \\
	0 & 0 & 0 & 0 & 3 & 3 & 0 & 0 \\
	0 & 0 & 0 & -1 & -1 & 0 & 0 & 0 \\
	0 & 0 & 0 & 0 & 0 & 0 & -1 & -1 \\
	0 & 0 & 0 & 2 & 1 & 0 & 1 & 1 \\
	\end{array}
	\right]}\,,\\
M_1&={\operatorname{diag}(2, 0, 2, 2, -6, 0, 2, 0)}\,,\\
M_2&={\operatorname{diag}(0, 0, 0, 0, 0, 0, 0, -2)}\,.
\end{align}

We obtain $\epsilon$-expansion of $\widetilde{\mathbf{J}}=\sum_n \epsilon^n \widetilde{\mathbf{J}}^{(n)}$ term-by-term using the formula
\begin{equation}
\widetilde{\mathbf{J}}^{(n+1)}=\int dx \left[\frac1{x}M_0+\frac{1}{x-1}M_1+\frac{1}{x+1}M_2\right]\widetilde{\mathbf{J}}^{(n)}+\text{const}
\end{equation}
and fixing the constant from small-$\beta$ asymptotics. In order to calculate this asymptotics, we use the method of expansion by regions \cite{BeneSmi1998}. The only nontrivial boundary conditions come from $O(\beta^{2\epsilon-1})$ term in small-$\beta$ asymptotics of  $J_1$ and $J_2$:
\begin{equation}
J_1\sim J_2\sim -\frac{2^{8 \epsilon -16} \pi ^{3 \epsilon -5} \Gamma (\epsilon )^2 \Gamma (2 \epsilon -1) \Gamma (3 \epsilon -2)}{\Gamma (4 \epsilon -1)} \beta ^{2 \epsilon -1}\,.
\end{equation}
As a result, the $\epsilon$-expansions of both $\widetilde{\mathbf{J}}$ and $\mathbf{J}$ is expressed in terms of HPLs. The expansions of $\mathbf{J}$ are presented in ancillary file.
Plugging the obtained expansions in the cross section expressed via $\widetilde{\mathbf{J}}$ we observe the cancellation of the terms $O(\epsilon^n)$ with $n=-4,\ldots,-1$. The $O(\epsilon^0)$ term gives us the result
\begin{multline}\label{eq:cs}
\sigma=\frac{(Z_1\alpha)^2(Z_2\alpha)^2}{\pi m^2}\bigg\{-\frac{1-\beta ^2}{12  \beta ^2}L^4+\frac{2
\left(23 \beta ^2-37\right) S_{3a}}{9   \beta ^2}+\frac{2 \left(11 \beta ^2-25\right) S_{3b}}{9   \beta ^2}-\frac{26 S_{2}}{9   \beta }\\
-\frac{\left(\beta ^6+217 \beta ^4-135 \beta ^2+45\right) L^2}{54   \beta ^6}
+\frac{5 \left(67 \beta ^4-48 \beta ^2+18\right) L}{27   \beta ^5}
-\frac{2 \left(78 \beta ^4-35 \beta ^2+15\right)}{9   \beta ^4}\bigg\}\,,
\end{multline}
\begin{align*}
S_{3a}&= \text{Li}_3\left(\frac{1-\beta}{1+\beta}\right)+L\, \text{Li}_2\left(\frac{1-\beta}{1+\beta}\right)-\frac{L^2}{2}  \log \left(\frac{2\beta }{1+\beta}\right)-\frac{L^3}{12}-\zeta_3\,,\\
S_{3b}&=\text{Li}_3\left(-\frac{1-\beta }{1+\beta}\right)+\frac{L}{2}\text{Li}_2\left(-\frac{1-\beta }{1+\beta}\right)+\frac{L^3}{24}-\frac{\pi ^2 L}{24}+\frac{3 \zeta_3}{4}\,,\\
S_{2}&=\text{Li}_2\left(-\frac{1-\beta }{1+\beta}\right)+L \log \left(\frac{\beta +1}2\right)-\frac{L^2}{4}+\frac{\pi ^2}{12}\,,\\
L&=\log\left(\frac{1+\beta}{1-\beta}\right)\,.
\end{align*}
\subsection{Asymptotics}
Given the expression \eqref{eq:cs}, it is easy to calculate both high-energy and low-energy asymptotics of the total cross section. For $\gamma\gg 1$ we have
\begin{multline}\label{eq:racah}
\sigma=\frac{(Z_1\alpha)^2(Z_2\alpha)^2}{\pi m^2}\bigg\{\frac{28 L_0^3}{27}-\frac{178
	L_0^2}{27}+\left(\frac{370}{27}+\frac{7 \pi ^2}{27}\right) L_0+\frac{7 \zeta_3}{9}-\frac{13 \pi ^2}{54}-\frac{116}{9}\\
-\frac{1}{\gamma ^2}\left[\frac{4 L_0^4}{3}-\frac{98 L_0^3}{27}+\frac{188 L_0^2}{27}-\left(\frac{172}{27}+\frac{25 \pi ^2}{54}\right) L_0-\frac{73 \zeta_3}{18}+\frac{5 \pi ^2}{27}+\frac{43}{27}\right]
+\ldots\bigg\}\,,
\end{multline}
where $L_0=\ln(2\gamma)$. The first line of Eq. \eqref{eq:racah} is the celebrated Racah's result \cite{Racah1937}, and the second line is the first correction to it. It is interesting to note that the correction is amplified by the fourth power of $L_0$.

For $\beta\ll1$ there is a strong compensation between separate terms in Eq. \eqref{eq:cs}, which leads to $\propto \beta^8$ suppression of the cross section. We have
\begin{equation}\label{eq:low}
\sigma=\frac{296(Z_1\alpha)^2(Z_2\alpha)^2 \beta^8}{55125\pi m^2}\left(1+\frac{7708 \beta ^2}{3663}+\ldots\right)\,.
\end{equation}
Recently, the small-$\beta$ asymptotics of the total cross section was discussed in Ref. \cite{Khriplovich2014}. The estimate $\sigma\propto \beta^5$ given there is in clear contradiction with our result \eqref{eq:low}. In fact, the estimate $\sigma\propto\beta^8$ can be justified in the following way. Using the kinematic constraints 
\begin{equation}
q_1\cdot u_1=q_2\cdot u_2=0\,,\quad (q_1+q_2)^2 > 4m^2
\end{equation}
for momentum transfers $q_{1,2}$, it is easy to understand that the main contribution to the cross section is given by the region where 
\begin{equation}
|\mathbf{q}_{1,2}|\sim m/\beta\,,\quad |\mathbf{q}_1+\mathbf{q}_2|\sim q_{1,2}^0\sim m\,.
\end{equation}
The characteristic momenta of the produced particles are of the order of their mass. Using these estimates, it is easy to count powers of $\beta$ in the total cross section. We have $\beta^{-3}$ from $d\mathbf{q}_1d\mathbf{q}_2$, $\beta^8$ from the photon propagators, $\beta^4$ from the denominator of electron propagator, and $\beta^{-1}$ from the flux of the colliding particles. As to the numerator of the electron propagator, one might check that it does not give $\beta^{-1}$ factor in the sum of two diagrams due to the estimate $\hat{u}_1\hat{q}_{1,2}\hat{u}_2-\hat{u}_2\hat{q}_{1,2}\hat{u}_1\sim O(\beta^0)$.

Using these estimates, we have derived the differential cross section at $\beta\ll 1$ and then obtained the leading term of \eqref{eq:low} by the direct integration. To avoid cluttering, we refrain from presenting fully differential cross section here. We only present the cross section, differential with respect to the energies of the produced particles:
\begin{gather}\label{eq:spectrum}
	\frac{d\sigma}{d\varepsilon _+d\varepsilon _-}\approx\frac{16 (Z_1\alpha)^2(Z_2\alpha)^2\beta ^8  p_- p_+}{45 \pi  \left(\varepsilon _-+\varepsilon _+\right){}^{10}}\left[(33 \varepsilon _+ \varepsilon _--49m^2)(\varepsilon _-^2+\varepsilon _+^2)-14 \varepsilon _+^2 \varepsilon _-^2+78 \varepsilon _+ \varepsilon _-m^2-32m^4\right]\,.
\end{gather}
Here $p_{\pm} =\sqrt{\varepsilon_\pm^2-m^2}$ and $\beta\ll 1$ is the relative velocity of the nuclei.
Integrating this cross section over $ \varepsilon_{\pm}$, we obtain the leading term of Eq. \eqref{eq:low}.

\section{Discussion and conclusion}
\begin{figure}
	\centering
	\setlength{\unitlength}{0.08\linewidth}
	\begin{picture}(0,0)
	\put(-0.25,1.4){\rotatebox{90}{$\sigma/\sigma_0$}}
	\put(5,1.5){\rotatebox{90}{$\delta$}}
	\put(2.5,-0.05){$u$}
	\put(7.8,-0.05){$u$}
	\end{picture}
	\includegraphics[width=10\unitlength]{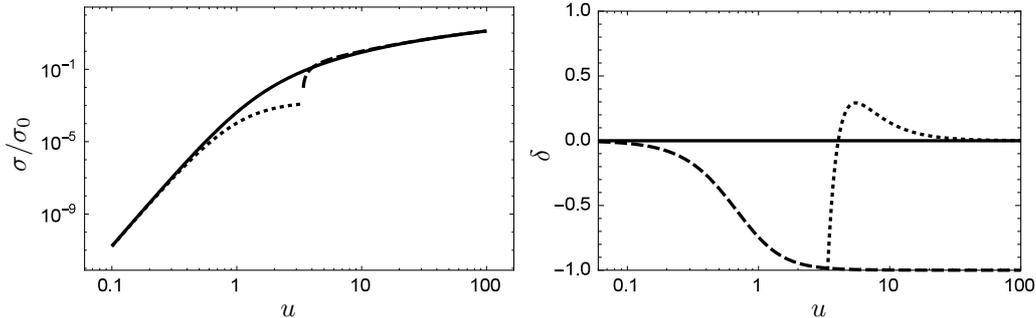}
	\caption{Left: cross section $\sigma$, Eq. \eqref{eq:cs}, in units of $\sigma_0=(Z_1\alpha)^2(Z_2\alpha)^2/m^2$ as a function of $u=\gamma\beta$ (solid curve). Dashed and dotted curves correspond to high- and low-energy asymptotics, respectively. Right: relative error $\delta=\sigma_{\text{h,l}}/\sigma-1$ of the high- and low-energy asymptotics.}
	\label{fig:cross_section}
\end{figure}

It is interesting to compare our result with the leading high-energy (Racah) and low-energy asymptotics $\sigma_{\text{h,l}}$. These asymptotics are given by the first line of Eq. \eqref{eq:racah} and the leading term of Eq. \eqref{eq:low}, respectively. Fig. \ref{fig:cross_section} demonstrates this comparison. One can see that both low- and high-energy asymptotics essentially depart from the exact result in the region $0.3\lesssim\gamma\beta\lesssim 10$.

Our approach based on the IBP reduction and calculation of the master integrals allows us, without additional efforts, to calculate the total Born cross section of the production of a pair of point-like scalar charged particles. We have
\begin{multline}
\sigma_s=\frac{\left(Z_1 \alpha \right)^2 \left(Z_2 \alpha \right)^2}{\pi  m^2} \bigg\{\frac{4 \left(\beta ^2-2\right) S_{3a}}{9 \beta ^2}+\frac{4 \left(\beta ^2-2\right) S_{3b}}{9 \beta ^2}-\frac{5 S_2}{9 \beta }\\
+\frac{\left(\beta ^6+\beta ^4-81 \beta ^2+45\right) L^2}{108 \beta ^6}-\frac{\left(10 \beta ^4-66 \beta ^2+45\right) L}{27 \beta ^5}-\frac{17 \beta ^2-15}{9 \beta ^4}\bigg\}\,.
\end{multline}
The high- and the low-energy asymptotics of this cross section have the form
\begin{equation}
\sigma_s=
\frac{\left(Z_1 \alpha \right)^2 \left(Z_2 \alpha \right)^2}{\pi  m^2} \begin{cases}
\frac{4 L_0^3}{27}-\frac{19 L_0^2}{27}+\frac{22+\pi ^2}{27} L_0+\frac{\zeta_3}{9}-\frac{5 \pi ^2}{108}-\frac{2}{9}+\ldots & \text{at }\gamma\gg 1\\
\frac{4 \beta ^4}{135 \pi } \left(1+\frac{27 \beta ^2}{35}+\frac{694 \beta ^4}{1225}+\ldots\right)& \text{at }\beta\ll 1
\end{cases}
\end{equation}
The leading term $L_0^3$ in the high-energy asymptotics agrees with the result obtained within the equivalent photon approximation.
Curiously, the low-energy asymptotics of the cross section for scalar particles scales differently ($\propto \beta^4$) than that for spinor particles. Inspection of the contributions of separate diagrams shows that in this limit only the contribution of seagull diagram survive.

It is interesting to discuss the applicability region of our results. Of course, by using the Born approximation we assume that $Z_{1,2}\alpha\ll 1$.
In principle, at small $\beta$ one may expect higher-order corrections of the relative order $Z_{1,2}\alpha/\beta$. However, since at small $\beta$ the velocities of the produced particles are not small (see, e.g., their spectrum \eqref{eq:spectrum}), we would guess that such corrections are forbidden. One may also expect corrections of the order $Z_1Z_2\alpha/\beta$ due to additional Coulomb exchanges between the nuclei, but they seem to be accompanied by the factor $m/M_{1,2}\ll1$, where $M$ are the masses of the nuclei. In fact a stronger condition $m/(\beta M_{1,2})\ll 1$ is definitely required because otherwise the limit $M_{1,2}\to \infty$ is no longer valid. Nevertheless, we must admit that the determination of the correct magnitude of the higher-order effects at small $\beta$ requires a separate examination.
\paragraph*{Acknowledgments} We are grateful to A.I. Milstein for the interest to our work and useful discussions. This work has been supported by Russian Science Foundation (Project No. 14-50-00080). Partial support of RFBR through Grant No. 15-02-07893 is also acknowledged.
\appendix
\bibliographystyle{elsarticle-num}

\end{document}